# Dirac semimetal strontium iridate thin films with strong spin-orbit interaction for magnetic heterostructures

Gennady A. Ovsyannikov[a*], Nikita V. Dubitskiy[a,b#], Georgi D. Ulev[a,c], Karen Y. Constantinian[a], Ivan E. Moskal[a], Victoria A. Baydikova[a], Andrei M. Petrzhik[a], Anton V. Shadrin[a,c], Alexei V. Mashirov[a]

[a]Kotelnikov Institute of Radio Engineering and Electronics, Russian Academy of Sciences, 125009, Moscow, Russia

[b]National Research University "Higher School of Economics", Faculty of Physics, 101000, Moscow, Russia.

[c]Moscow Institute of Physics and Technology (National Research University), Dolgoprudniy, Moscow Reg. 141701, Russia.

Corresponding authors: *gena@hitech.cplire.ru (G.A. Ovsyannikov)

[#]nikita.dubitskiy@gmail.com (Nikita V. Dubitskiy)

**Abstract**

The structural crystal features, electron transport and magnetotransport of the epitaxial strontium iridate ($SrIrO_3$) and iridate/manganite $SrIrO_3/La_{0.7}Sr_{0.3}MnO_3$ heterostructure have been investigated. The influence of epitaxial strain relaxation caused by the lattice mismatch of parameters of $SrIrO_3$ films and five substrates: $SrTiO_3$, $NdGaO_3$, $(LaAlO_3)_{0.3}(Sr_2TaAlO_6)_{0.7}$, $LaAlO_3$, and $Pb(Mg_{1/3}Nb_{2/3})O_3$-$PbTiO_3$ on electron and magnetic transport have been observed. A pronounced impact of strong spin-orbit interaction on characteristics of $SrIrO_3$ films have been revealed by means of X-ray photoelectron spectroscopy, magnetoresistance and Hall-resistance measurements at temperatures T = 2–300 K. These findings highlight the tunability of spin–orbit-driven transport phenomena in strain-controlled $SrIrO_3$-based epitaxial systems, relevant for future spintronic oxide heterostructures. The contribution of Kondo scattering on temperature dependence of $SrIrO_3$ films resistance was observed.

**Keywords**: strontium iridate, epitaxial strontium iridate films;; Dirac semimetal; spin–orbit coupling; magnetotransport; Kondo scattering; Hall effect

## 1. Introduction

Iridium oxides (iridates), transition metal oxides with 5d electron orbitals are promising materials for realizing non-trivial electronic states such as topological insulators, unconventional superconductors, and Dirac semimetals [1–5]. In particular, the Ruddlesden-Popper series of strontium iridates $Sr_{(n+1)}Ir_nO_{(3n+1)}$, exhibits an evolution from the three-dimensional correlated metal SrIrO$_3$ (n→∞) to the two-dimensional Mott insulator Sr$_2$IrO$_4$ (n=1) [1]. The insulating state arises from the crystal-field splitting of the 5d levels into e$_g$ and t$_{2g}$ manifolds, in which the partially filled t$_{2g}$ band further splits into $J_{eff}$ = 3/2 and $J_{eff}$ = 1/2 states due to the strong spin-orbit coupling associated with iridium ions. The Mott gap opens at $J_{eff}$ = 1/2 if the Coulomb interaction becomes comparable to the spin-orbit interaction.

Strontium iridate (SrIrO$_3$) possesses an orthorhombic GdFeO$_3$-type perovskite structure (space group *Pbnm*) characterized by strong spin–orbit interaction and electron correlations, giving rise to Dirac-semimetal-like behavior [5–10]. The lattice parameters of SrIrO$_3$ are $a$ = 0.56 nm, $b$ = 0.558 nm, and $c$ = 0.789 nm. This can be treated as a pseudo-cubic structure with $a \approx b \approx c \approx 0.396$ nm [11–14]. The spin-orbit interaction in iridates and their peculiar electronic density of states result in a large spin Hall effect, which plays a key role in spin current generation and its detection via the inverse spin Hall effect at ferromagnetic/normal metal interfaces [7–10].

The orthorhombic iridate SrIrO$_3$ is difficult to obtain in crystal form as it requires extremely high pressures to synthesize [13], only polycrystal perovskite SrIrO$_3$ were reported in [14] and no reports on bulk single-crystal perovskite SrIrO$_3$ to-date. In contrast, the orthorhombic phase readily forms in thin films at ambient pressure owing to dimensional and epitaxial stabilization effects [15]. SrIrO$_3$ thin films have recently attracted considerable attention because their heterostructures exhibit the topological Hall effect [16, 17], anomalous Hall effect [18], and robust metallic conductivity in ultrathin films [19].

In this work, we report on structural features and discuss relaxation of mechanical strain in thin epitaxial SrIrO$_3$ films grown by RF cathode sputtering on the substrates with pseudocubic lattice parameter of substrate from 0.379 nm up to 0.402 nm. It was found that due to the mismatch of the lattice parameters of the film and the substrate, significant compressive strain arises during the growth, leading to deformation of SrIrO$_3$ lattice structure. We present also results on magnetoresistances and Hall resistance of SrIrO$_3$ and La$_{0.7}$Sr$_{0.3}$MnO$_3$ epitaxial films and SrIrO$_3$/La$_{0.7}$Sr$_{0.3}$MnO$_3$ heterostructure.

## 2. Crystal structure of SrIrO$_3$ film
### 2.1 Fabrication technique

Epitaxial thin films of SrIrO$_3$ and La$_{0.7}$Sr$_{0.3}$MnO$_3$ with thicknesses ranging from 10 to 50 nm were grown on single-crystal substrates (110)NdGaO$_3$, (001)SrTiO$_3$, (110)(LaAlO$_3$)$_{0.3}$(Sr$_2$TaAlO$_6$)$_{0.7}$,(LSAT), LaAlO$_3$, and (110)Pb(Mg$_{1/3}$Nb$_{2/3}$)O$_3$-PbTiO$_3$ (PMN–PT) by RF magnetron sputtering at a substrate temperature of 770–800 °C in an Ar/O$_2$ gas mixture at a total pressure of 0.3–0.5 mbar [11, 20, 21]. The sputtering target was prepared from a pressed SrIrO$_3$ powder, followed by annealed at 1000 °C. X-ray diffraction confirmed the single-phase nature of the target. After the deposition, the films and heterostructures were cooled to room temperature in ambient pressure of oxygen.

## 2.2. X-ray diffraction data

The structural properties of $SrIrO_3$ films and $SrIrO_3/La_{0.7}Sr_{0.3}MnO_3$ heterostructures were investigated by X-ray diffraction using a Rigaku SmartLab diffractometer equipped with a rotating Cu anode. Measurements were performed in parallel-beam geometry with a Ge(220)×2 monochromator and $CuK_{\alpha 1}$ radiation ($\lambda$ = 1.54056 Å). Intense and well-defined (00k) reflections ($k$ = 1, 2, 3, 4) were observed for films grown on (001)$SrTiO_3$, (001)LSAT, and (110)$NdGaO_3$ substrates (Fig. 1), while for films deposited on (110)PMN-PT and (110)$LaAlO_3$ substrates, (nm0) $SrIrO_3$ reflections were detected (see Fig. S1 in the Supplementary Material).

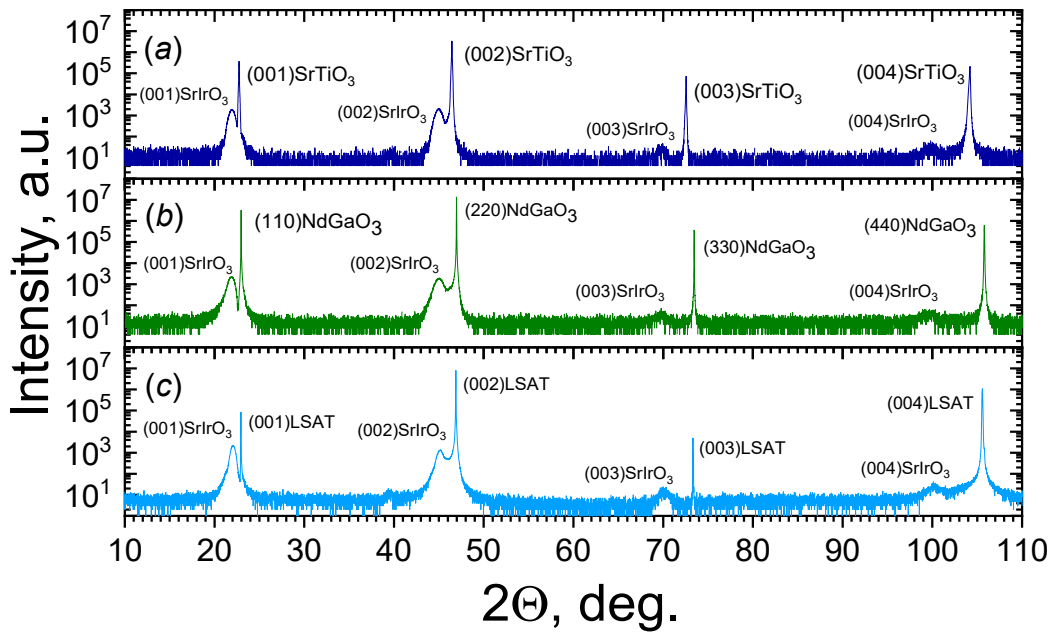

Fig. 1. XRD Bragg reflections for a) $SrIrO_3$ film grown on (001) $SrTiO_3$, b) $SrIrO_3$ film grown on (110)$NdGaO_3$ film, c) $SrIrO_3$ grown on (001) LSAT.

Analysis of the X-ray diffraction patterns (Fig. 1 and Fig. 1S) revealed a clear dependence of the interplanar spacing of $SrIrO_3$ films ($c_f$) on the lattice parameters of the substrates ($a$). The lattice mismatch ($m$) between $SrIrO_3$ crystal and the substrate, calculated as $m = c/a - 1$ (where $c$ denote the pseudo-cubic lattice parameter of the substrate) as estimated to be 1.5 % for $SrIrO_3/SrTiO_3$. The effective unit-cell volume of the $SrIrO_3/SrTiO_3$ film was $V = a^2 c_f$ = 0.0615 nm$^3$ compared with $V$ = 0.0621 nm$^3$ for bulk $SrIrO_3$ crystal. This lattice mismatch induces a biaxial compressive strain in the $SrIrO_3$ film during growth due to the influence of the substrate.

The films on (001)SrTiO$_3$, (001)LSAT, (110)NdGaO$_3$, and (110)LaAlO$_3$ substrates exhibit lattice compression (the mismatch parameter $m$ = 1.5–4.5 % is positive), which is accompanied by an increase in the interplanar spacing of the film unit cell to $c_f$ =0.402–0.404 nm [22]. A simultaneous decrease in the in-plane parameter determined by the substrate, to $a$ = 0.387–0.391 nm is also observed. For the SrIrO$_3$/LaAlO$_3$ film, the unit-cell volume is compressed to $V$ = 0.057 nm$^3$, which is smaller than that of the orthorhombic SrIrO$_3$ crystal ($V$ = 0.062 nm$^3$).

Of particular interest is the SrIrO$_3$ /SrTiO$_3$ film, which, despite a lattice mismatch of $m$ = 1.47%, exhibits a unit-cell volume of $V$ = 0.0613 nm$^3$, very close to that of bulk SrIrO$_3$. In contrast, for the SrIrO$_3$/PMN–PT film (Fig. 1S Supplementary Material), an in-plane tensile strain of $m$ = −1.5 % is observed for the substrate lattice parameter $a$ = 0.402 nm, leading to an increase in the effective unit-cell volume to $V$ = 0.0635 nm$^3$ compared with the bulk value of $V$ = 0.0621 nm$^3$ for the ideal single crystal [13]. The dependence of $V$ on $m$ is follows a nearly parabolic trend (see Fig. 2S).

Table 1. Structural and morphological parameters of the SrIrO$_3$ film. $c$ and $a$ denote the interplanar spacing of the SrIrO$_3$ film and the pseudocubic lattice parameter of the substrate, respectively. $m$ and $V$ represent the effective lattice strain and the unit-cell volume of the crystal, respectively. RMS corresponds to the root-mean square surface roughness of the film.

| Sample | $c$, nm | $a$, nm | $V$, 10$^3$ nm$^{-3}$ | $m$, % | RMS, nm |
|---|---|---|---|---|---|
| Crystal SrIrO$_3$ | 0.396 | 0.396 | 62.1 | – | – |
| SrIrO$_3$/ SrTiO$_3$ | 0.403 | 0.390 | 61.3 | 1.5 | 1.47 |
| SrIrO$_3$/ NdGaO$_3$ | 0.400 | 0.386 | 59.6 | 2.6 | 1.44 |
| SrIrO$_3$/LSAT | 0.402 | 0.387 | 60.2 | 2.3 | 0.31 |
| SrIrO$_3$/LAlO$_3$ | 0.397 | 0.379 | 57.0 | 4.5 | 5.90 |
| SrIrO$_3$/PMN-PT | 0.396 | 0.402 | 63.9 | -1.5 | 6.11 |

The epitaxial growth of the SrIrO$_3$ film was confirmed by φ-scan and transmission electron microscopy (TEM) [20, 21]. For SrIrO$_3$/SrTiO$_3$ film (similarly for SrIrO$_3$/LSAT), the following epitaxial relationships were observed: (001)SrIrO$_3$ ∥ (001)SrTiO$_3$ and [100]SrIrO$_3$ ∥ [100]SrTiO$_3$, indicating cube-on-cube growth. A more complex growth mode was found for SrIrO$_3$ films deposited on (110)LaAlO$_3$ and (110)PMN–PT substrates (see Supplementary Material).

### 2.3. Morphology of SrIrO$_3$ films

Atomic force microscopy (AFM) was employed to examine the surface morphology of SrIrO$_3$ thin films grown on various substrates. Surface topography images were acquired in tapping mode using a Solver PRO-M scanning probe microscope (NT-MDT). The microscope provides a vertical resolution of about 1 nm and enables examination of samples up to 12 × 12 mm$^2$ in size, with a lateral positioning accuracy of approximately 5 μm (see Fig. 2).

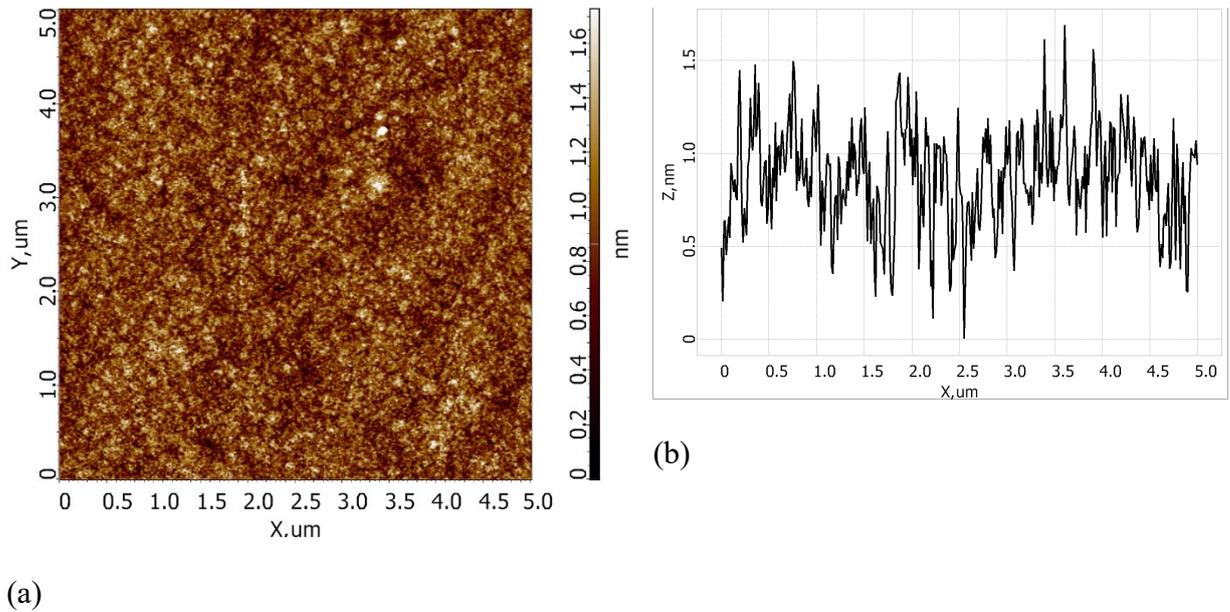

Fig. 2. (a) AFM image of the SrIrO$_3$/LSAT film. (b) Surface profile of the SrIrO$_3$/LSAT film measured along the horizontal line.

Figure 2(a) shows an AFM image of the SrIrO$_3$ film grown on the LSAT substrate (SrIrO$_3$/LSAT film). The 5 × 5 μm$^2$ topography scan reveals a granular structure morphology with a quasi-regular distribution of relief features. The SrIrO$_3$/LSAT film exhibited the lowest root-mean-square roughness (RMS) among all samples (0.31 nm); however, statistical analysis revealed anomalously high values of the skewness (11.00) and kurtosis (4.98) coefficients. These values indicate the presence of individual high protrusions on an otherwise smooth surface, possibly associated with substrate defects or specific features of the growth process. The height profile along the marked horizontal line (Fig. 2(b)) shows nanometer-scale variations, with a maximum peak-to-peak height of 0.7 nm.

Similar measurements were performed for SrIrO$_3$/NdGaO$_3$ and SrIrO$_3$/SrTiO$_3$. The SrIrO$_3$/NdGaO$_3$ film exhibited a higher RMS roughness (1.44 nm) compared with SrIrO$_3$/LSAT as well as a larger height range (25.58 nm) and kurtosis (4.98), indicating more pronounced local surface irregularities. SrIrO$_3$/PMN–PT and SrIrO$_3$/LaAlO$_3$ films showed the highest roughness values (RMS ≈ 6 nm ) and the largest height variations (up to 47.01 nm) among the studied films (see Supplementary Material).

These results demonstrate that the surface morphology of SrIrO$_3$ films strongly depends on the substrate [23]. The SrIrO$_3$/SrTiO$_3$ and SrIrO$_3$/NdGaO$_3$ films exhibit comparable roughness values, while the SrIrO$_3$/LSAT film shows a more uniform relief distribution. Despite its lowest RMS roughness, the SrIrO$_3$/LSAT film possesses a statistically non-uniform surface characterized by isolated protruding defects.

**2.4. X-ray photoemission spectroscopy**

The chemical composition and electronic structure of SrIrO$_3$ films were examined by X-ray photoelectron spectroscopy (XPS), which is based on the photoelectric effect induced by monochromatic X-ray radiation on the film. Shifts in the binding energy of photoelectron lines provide precise information on changes in the local environment of atoms. The measurements were carried out using a Theta Probe Spectrometer (Thermo Fisher Scientific, UK) at a residual gas pressure better than $1.3 \times 10^{-8}$ mbar. X-ray excitation was generated by the AlKα source with a photon energy of $E = 1486.6$ eV. The absolute energy resolution of the spectrometer, determined from the Ag 3d$_{5/2}$ core level, was 0.46 eV. The X-ray beam size was set to 400 μm, and the energy analyzer operated in the fixed analyzer transmission (FAT) mode. The absolute uncertainty in the measured photoelectron kinetic energy did not exceed 0.1 eV. The measurements were performed in two stages. In the first stage, survey spectra were recorded with an energy step of 1 eV and a pass energy of 200 eV (see Fig. 3(a)). In the second stage, high-resolution spectra of individual core levels corresponding to the Sr and Ir peaks were collected with an energy step of 0.100 eV and a pass energy of 50 eV (Fig. 3(b), (c)).

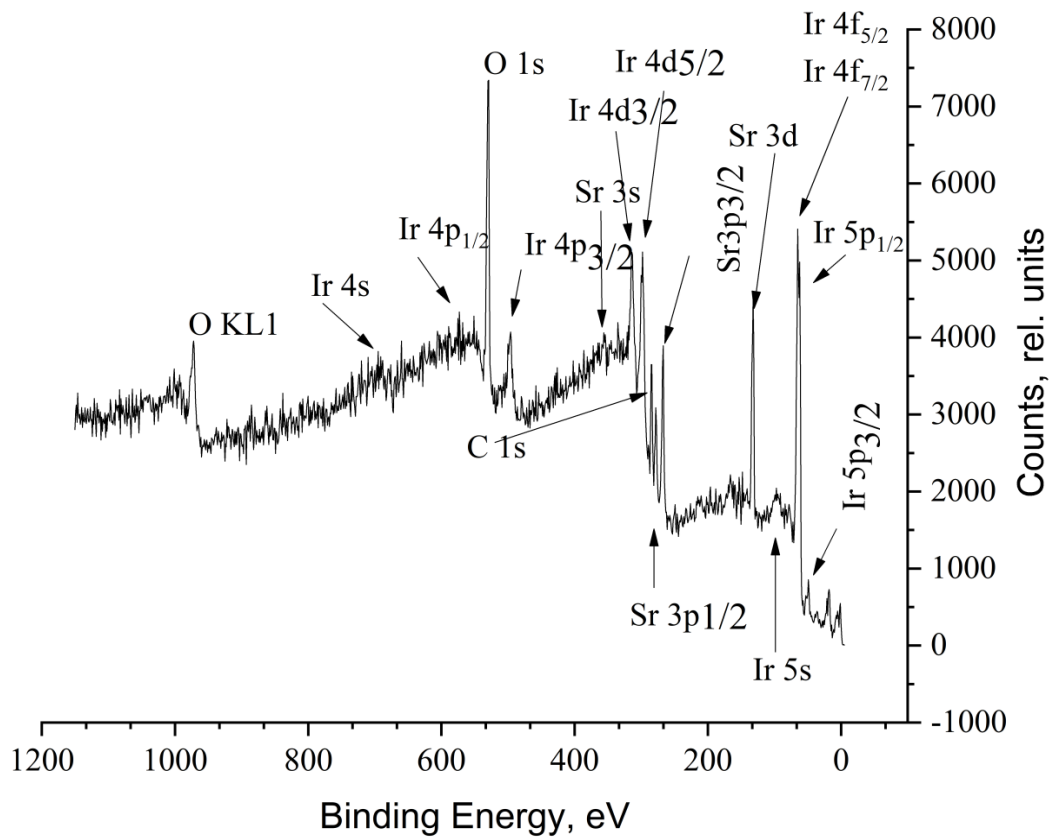

(a)

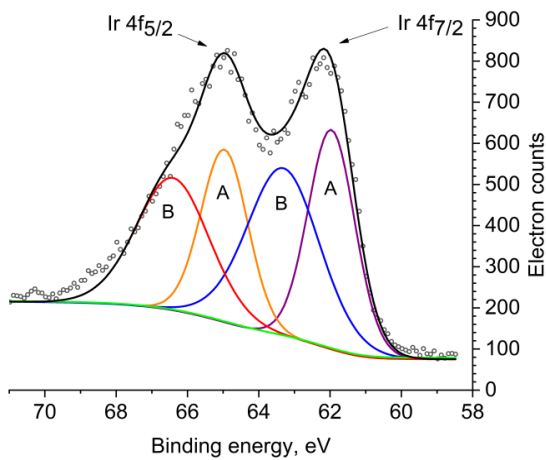

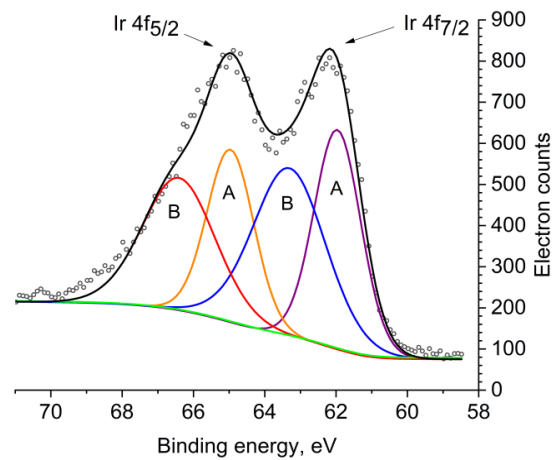

(b)                                          (c)

Fig. 3. (a) Survey XPS spectrum of the $SrIrO_3/SrTiO_3$ film. The binding-energy values for peak identification (element and sublevel assignment) were taken from the NIST XPS Database [26]. (b) High-resolution Ir 4f spectrum. The orange and red curves represent Voigt-function fits of the Ir $4f_{5/2}$ peak, while the violet Ir(A) $4f_{7/2}$ and blue Ir(B) $4f_{7/2}$ components correspond to the fitted Ir $4f_{7/2}$ peak. (c) Sr 3d spectrum: the red and blue curves correspond to the fitted Sr $3d_{5/2}$ and Sr $3d_{5/2}$ peaks, respectively. The green line indicates the Shirley-type background due to inelastic photoelectron scattering.

The overview XPS spectrum of the SrIrO$_3$/SrTiO$_3$ film is presented in Fig. 3(a). The Ir 4f core-level spectrum consists of two components, Ir 4f$_{7/2}$ and Ir 4f$_{5/2}$, originating from spin–orbit coupling (Fig. 3(b)). The spin–orbit splitting ($\epsilon_{SO}$) varies from 2.99 eV for the SrIrO$_3$/SrTiO$_3$ film to 3.10 eV for SrIrO$_3$/PMN–PT (Table 2). These values are consistent with the reported data for SrIrO$_3$ ($\epsilon_{SO} \approx 3.0$ eV) [24] and correspond to Ir$^{4+}$ oxidation states [25].

The magnitude of $\epsilon_{SO}$ is known to depend on the electronic structure and defect concentration. According to [25], $\epsilon_{SO}$ is particularly sensitive to the amount of oxygen vacancies in SrIrO$_3$ films. Films with higher $\epsilon_{SO}$ values generally exhibit fewer oxygen vacancies, indicating a composition closer to stoichiometry and a higher structural quality. Hence, the maximum splitting ($\epsilon_{SO} = 3.10$ eV) observed for SrIrO$_3$/PMN–PT films suggests its more optimal oxygen stoichiometry and lower concentration of magnetic impurities compared to the other films. For SrIrO$_3$/PMN-PT film may indicate its optimal oxygen stoichiometry and a lower concentration of magnetic impurities compared to other studied samples. The Sr 3d spectrum also reveals a typical doublet structure (Sr 3d$_{5/2}$ and Sr 3d$_{3/2}$) with a separation of about 1.8 eV, characteristic of Sr in an oxide environment (see Supplementary Material).

For a more detailed insight into the electronic structure, each of the Ir 4f$_{7/2}$ and Ir 4f$_{5/2}$ peaks was deconvoluted into two components, denoted Ir(A) and Ir(B), which correspond to Ir atoms in octahedra with different local distortions. The intensity ratio of the Ir(B)/Ir(A) varies from 1.14 to 1.35 depending on the substrate [25], implying different distributions of these states. The Sr/Ir atomic ratio ranges from 1.16 for SrIrO$_3$/SrTiO$_3$ to 1.37 for SrIrO$_3$/PMN–PT, correlating with the lattice mismatch and strain determined from X-ray diffraction analysis.

Table 2. XPS spectral parameters of epitaxial thin SrIrO$_3$ films.

| Sample | Sr, % | Ir, % | Sr/Ir % | Ir(B)/Ir(A) | $\epsilon_{SO}$. eV |
|---|---|---|---|---|---|
| SrIrO$_3$/SrTiO$_3$ | 53.81 | 46.19 | 1.16 | 1.20 | 3.03 |
| SrIrO$_3$/NdGaO$_3$ | 56.28 | 43.72 | 1.29 | 1.14 | 2.99 |
| SrIrO$_3$/LSAT | 56.94 | 43.06 | 1.32 | 1.35 | 2.99 |
| SrIrO$_3$/PMN–PT | 57.77 | 42.23 | 1.37 | 1.2 | 3.10 |

The SrIrO$_3$/LaAlO$_3$ film was etched with fluoride ions to clean the surface prior to the XPS measurements. However, this treatment modified the near-surface composition of the film, leading to a change in the Sr/Ir ratio. Therefore, the SrIrO$_3$/LaAlO$_3$ sample is not included in Table 2.

### 3. Electron and magnetic transport of the films and heterostructures
### 3.1. Electron transport of SrIrO$_3$ films

The electrical transport properties of SrIrO$_3$ films were measured using a four-probe method in the Montgomery configuration within temperature range of .2–300 K [27]. Fig. 4 presents the temperature dependence of the sheet resistance $R$(T) for SrIrO$_3$ films grown on five different substrates. The experimental $R$(T) data were analyzed by considering several contributions to the total resistance: the residual term $R_0$ associated with nonmagnetic impurities; the $qT^2$ and $pT^5$ terms corresponding to electron–electron and electron–phonon scattering mechanisms, respectively, the Kondo-scattering contribution $R_K$(T) [28, 29, 30–32]; and an additional term $c \cdot ln$(T/T$_0$) accounting for weak electron disorder effects, which becomes especially pronounced at low temperatures [33–35], as summarized in Eq. (1):

$$R(T) = R_0 + q\left(\frac{T}{T_1}\right)^2 + p\left(\frac{T}{T_2}\right)^5 + R_K\left(\frac{T_K}{T^2 + T_K^2}\right)^S + c \cdot \ln\left(\frac{T_3}{T}\right), \qquad (1)$$

where $s = 0.225$, and $T_K$ is the Kondo temperature [28−32]. This model fits the experimental data reasonably well (see Fig. 5S Supplementary Material). The Kondo-like resistivity term $R_K$, which scales with the concentration of magnetic impurities, could be related to oxygen vacancies [25]. *Ab initio* calculations for strained perovskite CaMnO$_3$ also support this assumption, showing that tensile strain may promote the formation of oxygen vacancies [24, 25].

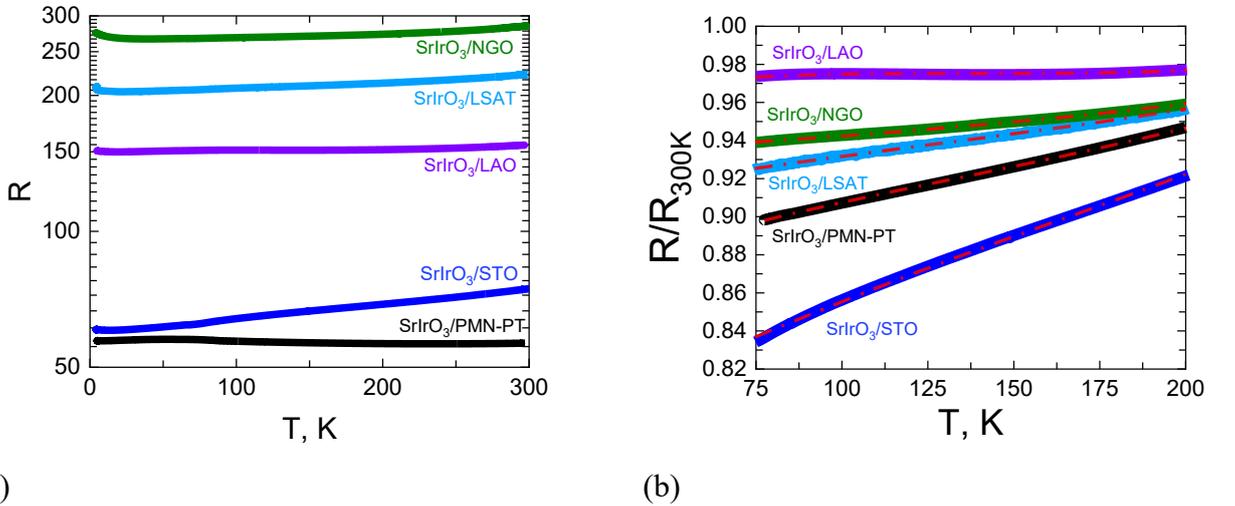

(a) (b)

Fig. 4. (a) Temperature dependence of the sheet resistance $R$(T) of SrIrO$_3$ films grown on NdGaO$_3$, SrTiO$_3$, LSAT, LaAlO$_3$, and PMN–PT substrates. (b) Normalized $R$(T) dependencies. Solid lines represent the fits according to Eq. (1).

Table 3 presents the electrical parameters for all films. It can be seen that the Kondo temperature $T_K$ varies significantly for films grown on different substrates, ranging from 135 K for SrIrO$_3$/PMN–PT to 4.8 K for SrIrO$_3$/SrTIO$_3$. The relatively high $T_K$ value for the latter may be attributed to deviations from the perturbative limit of the Kondo model [29, 30]. The strongest effect of magnetic impurities (maximum value $T_K = 135$ K) is observed in films for SrIrO$_3$/PMN–PT ($m = -1.5\%$), but maximum value $R_K = 127$ Ω is observed for SrIrO$_3$/LSAT film (m = 2.3%) According to Table 3, the electron–phonon contribution $pT^5$ and the electron–electron contribution $qT^2$ in Eq. (1) are negligible compared with the other terms. The parameter $q$ ranges from −0.09 to 1.85 mΩ and $p$ from −3 to 3 pΩ, whereas $R_0$ ranges from 43 to 238 Ω, $R_K$ from 18 to 127 Ω, and $c$ from 0.31 to 9.65 Ω. Thus the low-temperature resistivity is governed primarily by Kondo scattering and electron-electron disorder. The substrate dependence is systematic: $T_K$ ranges from 1.4 to 4.8 K for SrIrO$_3$/SrTiO$_3$, SrIrO$_3$/NdGaO$_3$, SrIrO$_3$/LSAT and SrIrO$_3$/LaAlO$_3$, while it reaches 135 K for SrIrO$_3$/PMN–PT, together with corresponding changes in $R_K$ and $c$. These trends indicate that epitaxial strain and the associated defect chemistry modulate the electronic structure and control charge transport in SrIrO$_3$ films.

Table 3. Approximation parameters of the $R$(T) dependence for SrIrO$_3$ films grown on SrTiO$_3$, NdGaO$_3$, LSAT, LaAlO$_3$, and PMN–PT substrates, obtained using Eq. (1).

| Film | $T_K$, K | $R_K$, Ω | $R_0$, Ω | $q$, mΩ | $p$, pΩ | $c$, Ω | $T_1$, K | $T_2$, K | $T_3$, K |
|---|---|---|---|---|---|---|---|---|---|
| SrIrO$_3$/SrTiO$_3$ | 4.8 | 47 | 43 | 0.35 | 3 | 7.38 | 2.2 | 13.4 | 21.2 |
| SrIrO$_3$/NdGaO$_3$ | 3.9 | 83 | 238 | 0.01 | 1 | 9.65 | 9.2 | 0.7 | 18.8 |
| SrIrO$_3$/LSAT | 1.4 | 127 | 192 | 0.36 | 3 | 4.9 | 1.6 | 12.3 | 20.2 |
| SrIrO$_3$/LAlO$_3$ | 2.2 | 18 | 144 | -0.09 | 2 | 2.81 | 1.4 | 1 | 15.8 |
| SrIrO$_3$/PMN–PT | 135 | 40 | 45 | 1.85 | -3 | 0.31 | 8.1 | 44.4 | 116.9 |

**3.2. Magneto- and Hall- resistance of SrIrO$_3$ films and the magnetic heterostructure**

Nowadays a transfer of the spin angular momentum through the functional interfaces in magnetic heterostructures is viewed as a promising tool for development of spintronic devices. A pure spin current can be induced by spin pumping in a ferromagnet/metal heterostructure assuming to exploit the spin-orbit interaction in N-layer. SrIrO$_3$ film for the "metal" layer and La$_{0.7}$Sr$_{0.3}$MnO$_3$ film are good candidates for the heterostructure [36-38]. SrIrO$_3$/La$_{0.7}$Sr$_{0.3}$MnO$_3$ heterostructures were fabricated by sequential deposition of La$_{0.7}$Sr$_{0.3}$MnO$_3$ and SrIrO$_3$ films *in situ* [20, 21].

Electron transport and magnetoresistance of SrIrO$_3$ films and SrIrO$_3$/La$_{0.7}$Sr$_{0.3}$MnO$_3$ heterostructure were studied in Hall-geometry configuration (see Fig. 5). The directions of applied electric current I and external magnetic field H are shown in the inset of Fig. 5. This scheme could be used as well for evaluation of spin-Hall angle $\theta_{SH}$ and spin magnetoresistance studies by means of changing angle $\varphi$.

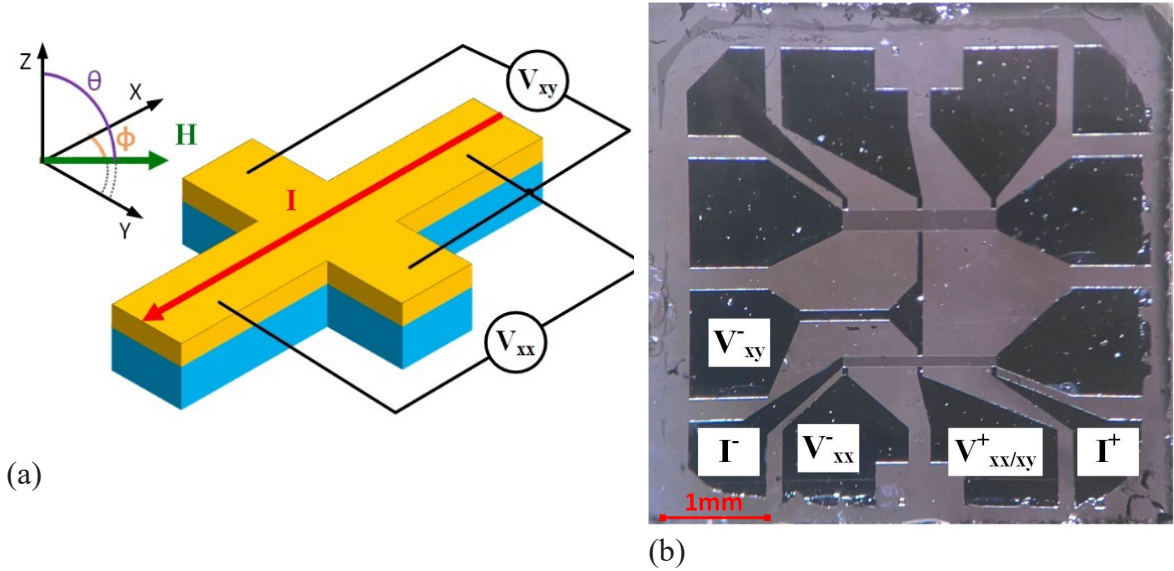

(a)

(b)

Fig. 5. (a) Scheme of magnetoresistive measurements. The DC current I directed along the X-axis, voltages $V_{XX}$ and $V_{XY}$ are measured for longitudinal (Ohmic) $R_{xx}$ and the transverse (Hall) $R_{xy}$ resistances. The inset on the left shows the directions of magnetic field H; (b) photo of the substrate with the patterned 2 thin film bars with width either $W$ = 0.1 mm, or 0.5 mm. Black parts correspond to Pt contacts marked by labels.

The experimental samples were patterned in form of "Hall-bars" with width either $W$ = 0.1 mm, or 0.5 mm and length $L$ = 1.3 mm (see Figure. 5b with labels for the bar with $W$ = 0.1 mm) using photolithography and ion etching. Resistance $R(T)$ = V(T)/I of thin films and heterostructure were studied at $T$ = 2–300 K using Keithley nanovoltmeters and current source (Keathley 2600), the sample temperature was measured by Lake Shore sensor [39, 40]. The contact pads (Fig.5b) were located on the top of either the films $SrIrO_3$, $La_{2/3}Sr_{1/3}MnO_3$ or heterostructure $SrIrO_3//La_{2/3}Sr_{1/3}MnO_3$.

Figure 6 shows the temperature dependences of the square film resistance $R_\square = R_{XX} W/L$ for $SrIrO_3$ film (thickness $d_{SIO}$ = 35 nm), $La_{0.7}Sr_{0.3}MnO_3$ ($d_{LSMO}$ = 40 nm) and the heterostructure $SrIrO_3/La_{0.7}Sr_{0.3}MnO_3$ ($d_{SIO}$ = 10 nm, $d_{LSMO}$ = 25 nm). A power-low temperature dependence of resistance $R(T) \sim T^{5\,2}$ for $La_{0.7}Sr_{0.3}MnO_3$ is inherent for half-metal ferromagnets (in particular for manganites) as shown by the fitting curve in Fig.6a for temperature range $T$ = 50–250 K[46]. With cooling the difference in resistances between $SrIrO_3$ and $La_{0.7}Sr_{0.3/3}MnO_3$ increases. The resistively shunting by $La_{0.7}Sr_{0.3/3}MnO_3$ layer affects also the heterostructure. At low temperatures $T < 20$ K an increase of resistance, caused also by an impact of Coulomb scattering and localization effects took place for all three structures [41,42]. Fig.6b shows resistance rise at $T < 20$ K for $SrIrO_3$ film on $SrTiO_3$ and $NdGaO_3$ substrates, while Fig.6c demonstrates similar behavior for $La_{2/3}Sr_{1/3}MnO_3$ and $SrIrO_3/La_{2/3}Sr_{1/3}MnO_3$. In the both cases the thickness of $La_{0.7}Sr_{0.3}MnO_3/NdGaO_3$ film exceeds the critical thickness of dead layer (2–3 nm) where ferromagnetism could be suppressed or granular structure realized [42].

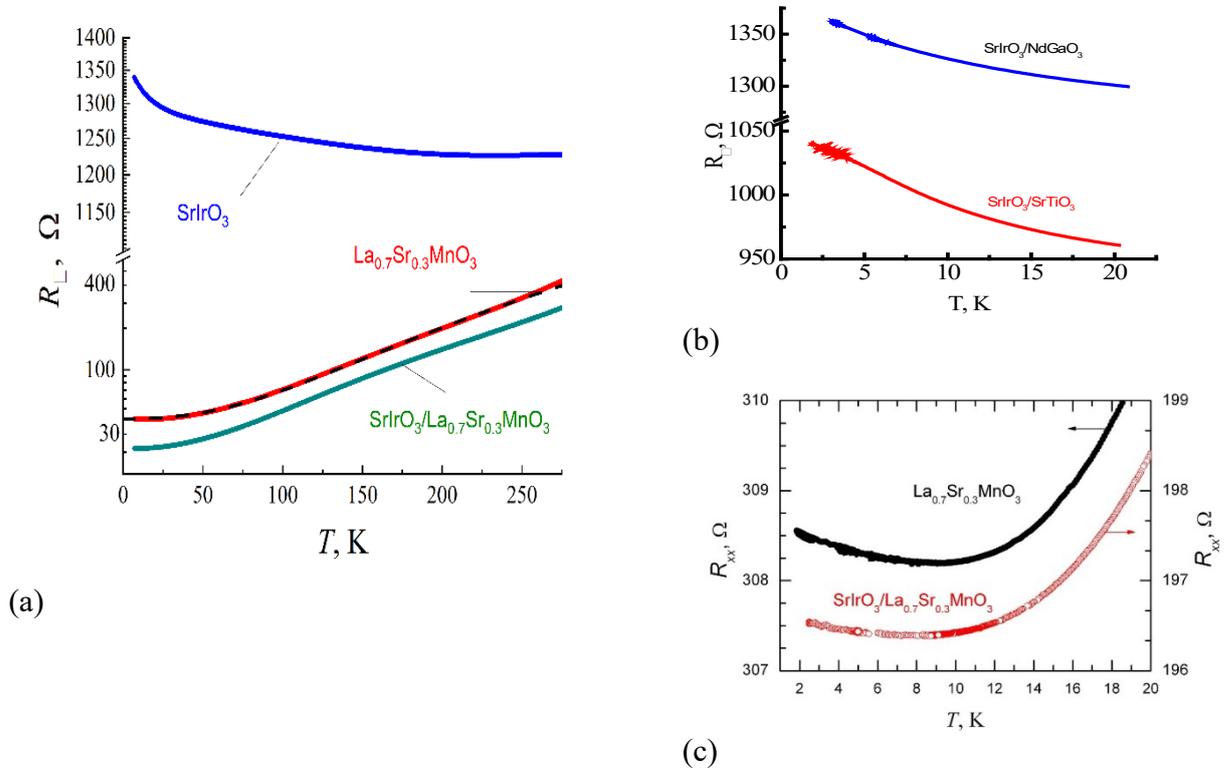

Fig. 6. (a) Temperature dependences of resistance $R_\square$ for either thin films SrIrO$_3$, La$_{0.7}$Sr$_{0.3}$MnO$_3$ and heterostructure SrIrO$_3$/La$_{0.7}$Sr$_{0.3}$MnO$_3$ deposited on NdGaO$_3$ substrate at H = 0. Black dashed line shows power-low $\propto T^P$ ($p$ = 5/2) theoretical approximation. (b) Temperature dependences of the resistance of both SrIrO$_3$/SrTiO$_3$ and SrIrO$_3$/ NdGaO$_3$ films, (c) Temperature dependences of $R_{XX}$ for La$_{0.7}$Sr$_{0.3}$MnO$_3$/NdGaO$_3$ film and SrIrO$_3$/La$_{0.7}$Sr$_{0.3}$MnO$_3$ heterostructure on NdGaO$_3$ substrate.

Note, sharp $R_{XX}$ rise at $T$ < 5 K influenced on $R_{XX}(T)$ for heterostructure, shifting the minimum position by 2 K from $T_{min}$ of La$_{0.7}$Sr$_{0.3}$MnO$_3$ was reported recently in [43].

A change in a sign altering of magnetoresistance and an anomalous Hall effect response in manganites were observed in the experiment [44-47]. Figure 7a shows normalized by R(0) dependences of ΔR = R(H) − R(0) on the magnetic field H (MR). Whereas the SrIrO$_3$ has negative MR, the sign of MR has changed twice both in La$_{0.7}$Sr$_{0.3}$MnO$_3$ film and SrIrO$_3$/La$_{0.7}$Sr$_{0.3}$MnO$_3$ heterostructure at H < 1 T. Fig. 7a shows well-coinciding double tracks of magnetoresistance registration. A similar change in the magnetoresistance sign was reported in [48] for the ferromagnetic superlattice SrRuO$_3$/La$_{0.7}$Sr$_{0.3}$MnO$_3$. The authors of [48] attributed the appearance of the positive magnetoresistance sign to a weak antilocalization (WL) due to spin-orbit coupling, which manifests itself varying the thicknesses of the ferromagnets in the superlattice and magnetic anisotropy due to influence of $SrRuO_3$ layers in the superlattice grown on $SrTiO_3$. In our case, the positive sign of MR at H = 1.1–5.5 kOe on the $La_{0.7}Sr_{0.3}MnO_3$ on $NdGaO_3$ substrate hardly could be associated with suppression of WL. On the contrary, a coating $La_{0.7}Sr_{0.3}MnO_3$ by $SrIrO_3$ film with strong spin-orbit interaction (SOI) only reduces the MR difference about 1.5 times. Almost the same amount 1.57 is obtained for the ratio of $R_{min}$ values for $La_{0.7}Sr_{0.3}MnO_3$ film and the heterostructure.

Fig. 7b) shows the dependences of the Hall voltage at T~10 K when the electrical conductivity of $La_{0.7}Sr_{0.3}MnO_3$ can be considered as a metallic. The Hall response $V_{xy}(H)$ increases with H, demonstrating negative magnetoresistance at H > 1 T. From Hall resistance measurement we evaluated an effective carrier density (Eq. (2)):

$$R_H = \frac{V_{xy}d}{IB} = \frac{1}{|e|}\frac{(n_h\mu_h^2 - n_e\mu_e^2)}{(n_h\mu_h - n_e\mu_e)^2}, \quad (2)$$

where $n_e$, $n_h$, $\mu_e$, $\mu_h$ are densities and nobilities for electrons and holes, correspondingly. However, it is worth to note a greater mobility of electrons compared to holes $\mu_e > \mu_h$ [46, 47].

It is evident from Fig. 7a that the voltages $V_{xy}$ for all three samples (each with thickness $d_i$) depend linearly on H-field and the Hall resistance $R_H = V_{xy}d_i/\mu_0 H$ with a simplified estimation for effective carrier concentration $n_{eff} = 1/eR_H$, ($\mu_0$ and e are physical constants), Eq. (3):

$$n_{eff} = \frac{(n_h\mu_h + n_e\mu_e)^2}{(n_h\mu_h^2 - n_e\mu_e^2)} \quad (3)$$

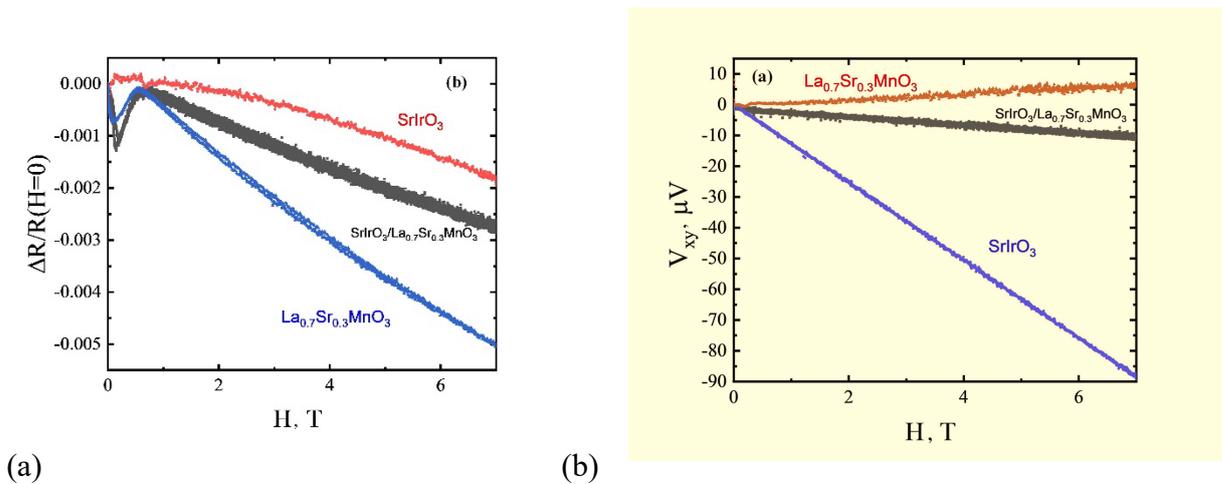

Fig. 7. (a) Magnetoresistance and (b) Hall voltage for $SrIrO_3$, $La_{0.7}Sr_{0.3}MnO_3$ films and $SrIrO_3$/ $La_{0.7}Sr_{0.3}MnO_3$ heterostructure measured at T = 10 K.

We obtain from data in Fig.7b at T=10 K for La$_{0.7}$Sr$_{0.3}$MnO$_3$ the hole type n$_{eff}$= 1.6 10$^{22}$ cm$^{-3}$, and for SrIrO$_3$ and SrIrO$_3$/La$_{0.7}$Sr$_{0.3}$MnO$_3$ the electron type n$_{eff}$= 1.4 10$^{21}$ cm$^{-3}$ and 1.25 10$^{22}$ cm$^{-3}$, respectively. Note, electron and holes mobilities ($\mu_e$, $\mu_h$) and concentrations (n$_e$, n$_h$) for thin SrIrO$_3$ films on SrTiO$_3$ substrate were accounted in [47] and for La$_{0.7}$Sr$_{0.3}$MnO$_3$ films in 44] giving a ratio of $\mu_e/\mu_h$ ~ 0.3 in mobility indicating a difference in effective masses of electrons and holes in these materials.

## 4. Conclusion

Epitaxial SrIrO$_3$ films were grown by RF magnetron sputtering on five single-crystal substrates. X-ray diffraction established biaxial strain (compressive or tensile) arising from lattice mismatch, accompanied by a systematic reduction of the effective unit-cell volume with increasing mismatch. XPS of the Ir 4f core level revealed spin-orbit splitting ≈3 eV for all films, consistent with Ir$^{4+}$ and indicating modest sample-to-sample variations that correlate with oxygen-vacancy content and structural quality.

The temperature dependence of the resistivity, analyzed over 4.2–300 K, is captured by a model including a residual term, electron–electron and electron–phonon contributions, a Kondo term, and a weak electron-electron disorder term $c \cdot \ln(T/T_0)$. At low temperatures the upturn is governed primarily by Kondo scattering and weak electron-electron disorder whereas electron–phonon and electron–electron terms are comparatively small. Hall measurements show that SrIrO$_3$ exhibits *n*-type conductivity and negative magnetoresistance, while La$_{0.7}$Sr$_{0.3}$MnO$_3$ behaves as a hole-type ferromagnetic semimetal. In SrIrO$_3$/La$_{0.7}$Sr$_{0.3}$MnO$_3$ heterostructures, the interface with strong spin–orbit coupling modulates the magnetoresistance and Hall response, consistent with resistive shunting and strain-driven changes to defect chemistry.

Overall, the results demonstrate that epitaxial strain and oxygen stoichiometry jointly control the electronic structure and charge transport in SrIrO$_3$ films and their manganite-based heterostructures. These insights provide a clear route for engineering spin–orbit-coupled oxide interfaces through substrate choice, strain state, and oxygen control, enabling the tuning of magneto- and Hall-transport functionalities relevant to oxide spintronics.

**Acknowledgements**

The authors are grateful to Yu.V.Kislinskii, A.A. Klimov, K.E. Nagornich, for experimental help and useful discussions. This work was supported by the Russian Science Foundation, project No. 23-49-00010.

**Supplementary Material**

1. Additional XRD data for the SrIrO3 film deposited on (110)-oriented substrates.

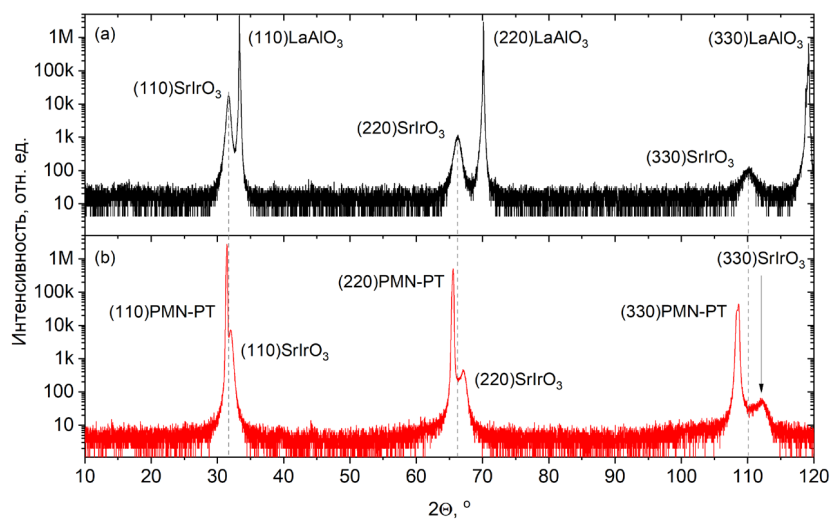

Fig. 1S. X-ray 2Θ/ω symmetric scan for SrIrO$_3$ films grown on substrates: (a) (110)LaAlO$_3$, (b) (110)PMN-PT.

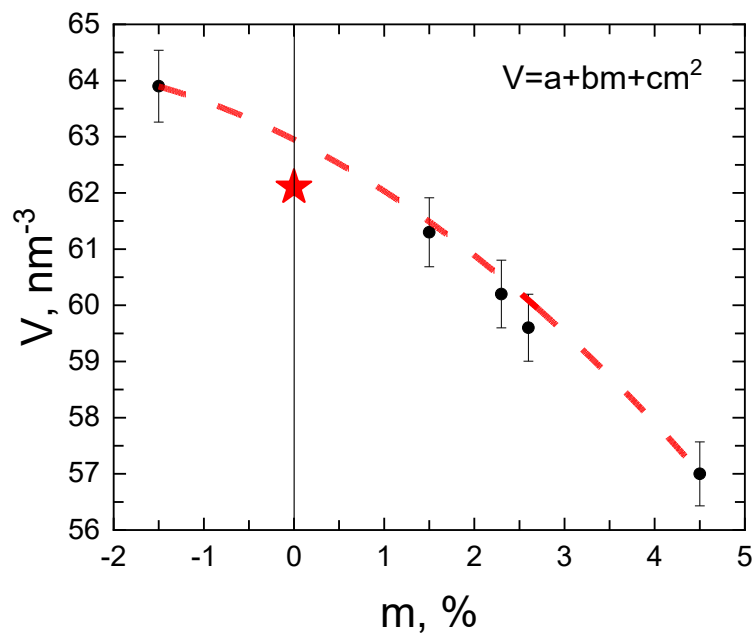

Fig. 2S. Dependence of the SrIrO$_3$ unit-cell volume on the effective strain parameter *m*.

3. Atomic force microscopy data for SrIrO$_3$/SrTiO$_3$ and SrIrO$_3$/NdGaO$_3$ films.

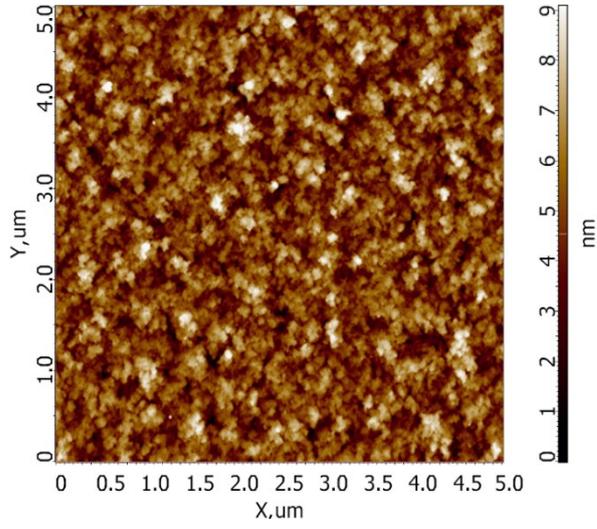

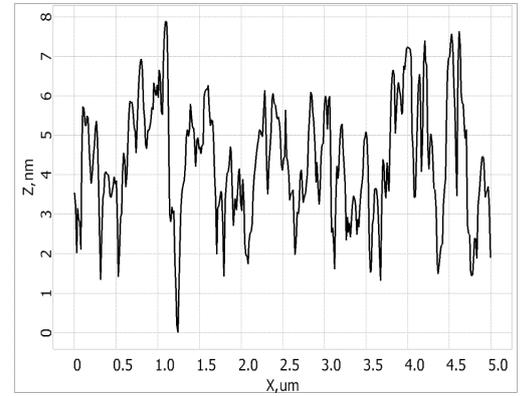

(a)

(b)

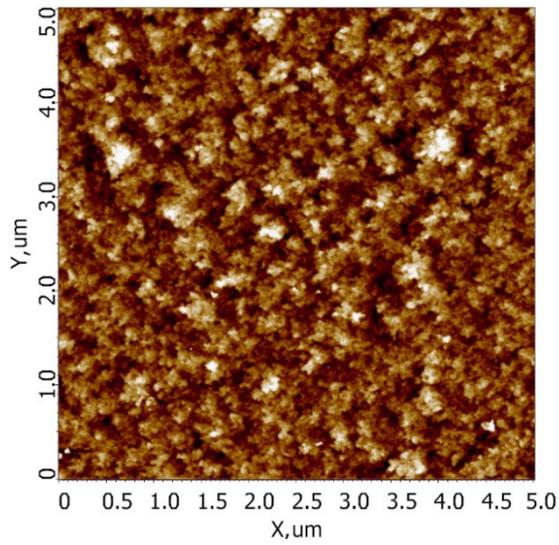

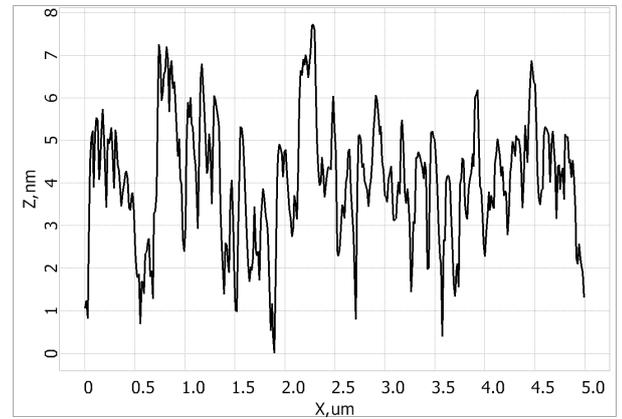

(c)

(d)

Fig. 3S. (a) AFM image of the surface of a SrIrO$_3$/SrTiO$_3$ film; (b) surface profile taken along one of the lines in the AFM image; (c, d) AFM images of SrIrO$_3$ films grown on NdGaO$_3$ substrates.

4. XPS spectra of SrIrO$_3$ films grown on different substrates.

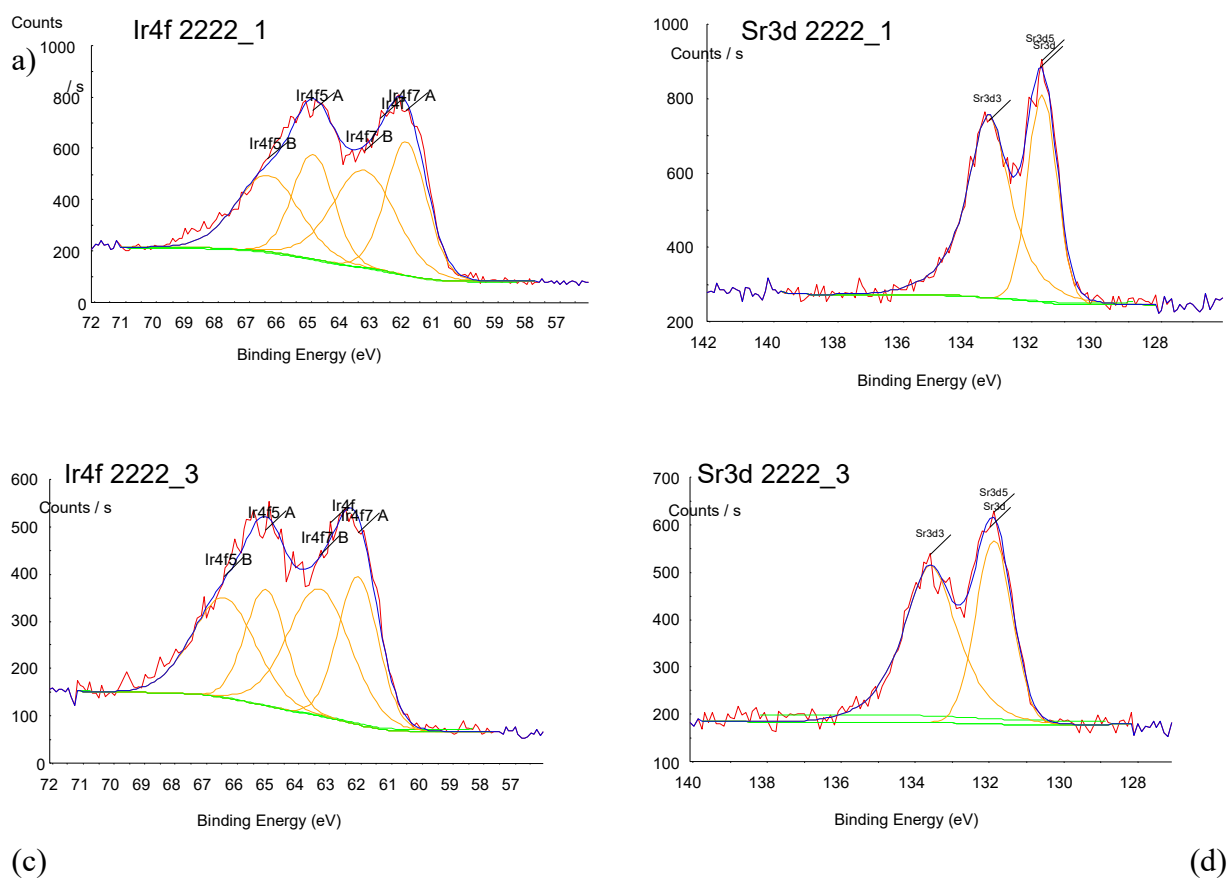

(c)                                                              (d)

Fig. 4S. XPS spectra of SrIrO$_3$ films: (a, b) SrIrO$_3$/NdGaO$_3$ and (c, d) SrIrO$_3$/LSAT. Panels (a) and (c) show the Ir 4f core-level spectra, and panels (b) and (d) correspond to the Sr 3d lines, respectively.

5. Fitting of the *R*(T) dependences for SrIrO$_3$ films grown on three substrates.

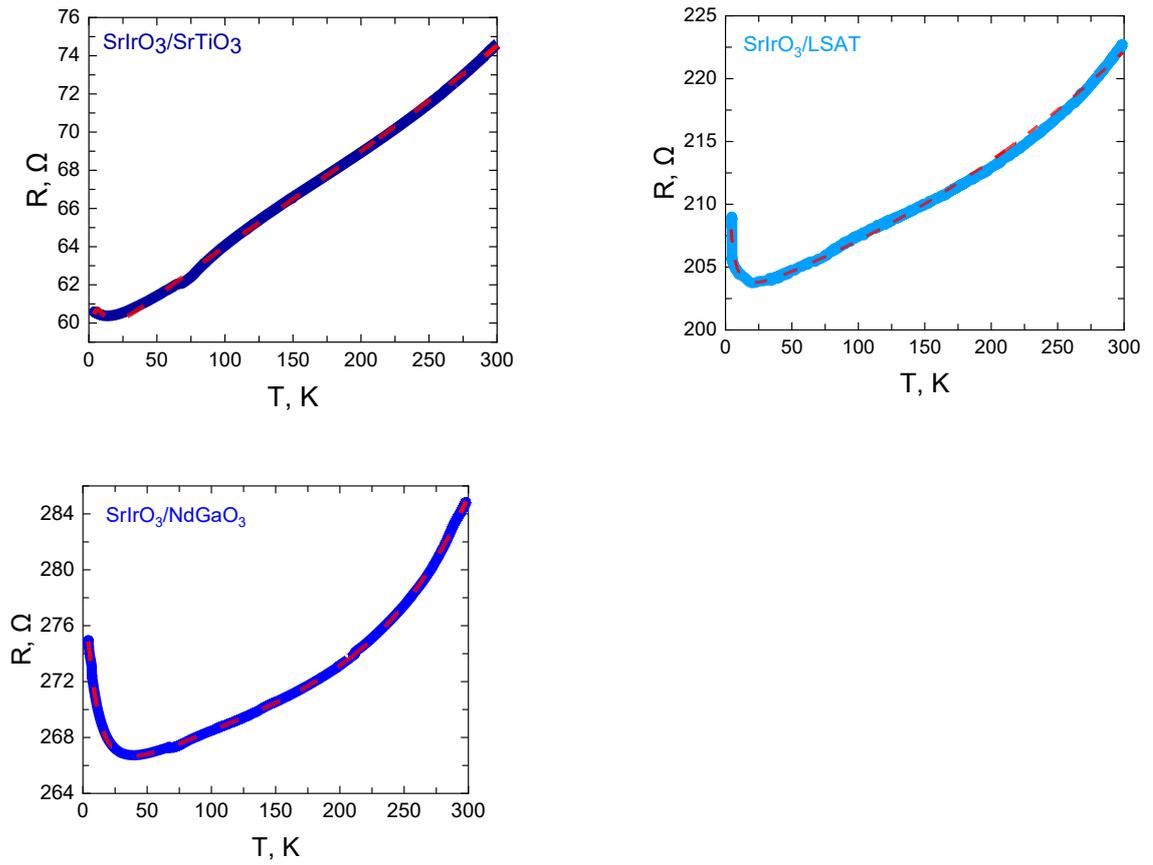

Fig. 5S. Temperature dependence of the sheet resistance $R_\square(T)$ for SrIrO$_3$ thin films grown on SrTiO$_3$, LSAT, and NdGaO$_3$ substrates. Red curves represent the fits to the experimental data according to Eq. (1).